%% file: intermit.tex
\documentstyle[seceq,supplement,epsf,amssymb]{ptptex}

\input{def}

\title{
Cycle expansions for intermittent maps
}

\author{
Roberto {\sc Artuso}$^{\dagger}
$\footnote{Roberto.Artuso@uninsubria.it,
Predrag.Cvitanovic@physics.gatech.edu,\\
Gregor.Tanner@nottingham.ac.uk
},
Predrag {\sc Cvitanovi\'c} $^{\dagger\dagger}$
and Gregor {\sc Tanner} $^{\dagger\dagger\dagger}$
\footnote{on leave from School of Mathematical Sciences, University of Nottingham, UK}
}

\inst{
$^{\dagger}$Dipartimento di Scienze Chimiche, Fisiche e Matematiche, Universit\`a
dell'Insubria and I.N.F.M., Sezione di Como, Via Valleggio 11, 22100 Como, Italy\\
$^{\dagger\dagger}$Center for Nonlinear Science,
School of Physics, Georgia Institute of Technology,
       Atlanta, GA 30332-0430, USA\\
$^{\dagger\dagger\dagger}$
   Quantum Information Processing Group,
   Hewlett-Packard Laboratories, Bristol BS34 8QP, UK}

\versiondate{
\today: submitted to Prog.~Theor.~Phys.}
\abst{
In a generic dynamical system
chaos and regular motion coexist side by side, in
different parts of the phase space. The border between
these, where trajectories are neither unstable nor stable but
of marginal stability, manifests itself through  intermittency,
dynamics where long periods of nearly regular motions are interrupted
by irregular chaotic bursts.
We discuss the \FPoper\ formalism for such systems, and
show by means of a 1-dimensional intermittent map
that intermittency induces branch cuts in \dzeta s.
Marginality leads to long-time
dynamical correlations, in contrast to the
exponentially fast decorrelations of purely chaotic dynamics.
We apply the periodic orbit theory to quantitative characterization
of the associated power-law decays.
}

\begin{document}

\maketitle

\section{Introduction}
\label{sec:intro}

In fluid dynamics one often observes long periods
of regular dynamics (laminar phases) interrupted by
irregular chaotic bursts, with the distribution of laminar phase intervals
well described by a power law.
This phenomenon is called {\em intermittency}\cite{MP79,intPM}, and it
is a very general aspect of dynamics, a shadow
cast by non-hyperbolic and  marginally stable phase
space regions.  Complete hyperbolicity
is indeed the exception rather than the rule:
almost any dynamical system of interest
exhibits a mixed phase space
where islands of stability or regular regions coexist with
hyperbolic regions with dynamics mixing exponentially fast.
The trajectories on the border between chaos and regular dynamics
are marginally stable, and trajectories from chaotic regions which come close
to marginally stable regions can stay `glued' there for
arbitrarily long times.  These intervals of regular motion
are interrupted by irregular bursts as the trajectory
is re-injected into the chaotic part of the phase space. How
the trajectories are precisely `glued' to
the marginally stable region is often hard to describe.
What coarsely looks like a border of an island in 
a Hamiltonian system with a mixed phase space, will under
magnification dissolve into infinities of island chains of
decreasing sizes and cantori \cite{ham}.

The existence of marginal or nearly marginal
orbits is due to incomplete
intersections of stable and unstable manifolds in a Smale horseshoe type
dynamics. Following the stretching and folding of the
invariant manifolds in time one will inevitably find phase space points at
which the stable and unstable manifolds are almost or exactly tangential
to each other, implying non-exponential separation of nearby
points in phase space or, in other words,  marginal stability.

How to deal with the full complexity of a typical Hamiltonian system
with a mixed phase space is a very difficult and still open
problem. Nevertheless, it is possible to learn quite a bit about
intermittency by considering rather simple examples. Here we
shall restrict our considerations to 1-dimensional maps
which in the neighborhood of a single marginally stable fixed point at
$x$=0 take the form
\beq
x\mapsto f(x) \sim x+c x^{1+s}
\ee{intereq:mapprot}
for $x \to 0$ and are expanding everywhere else.
Such a map may allow for escape or the dynamics may be bounded, like the
Farey map
\beq
x \mapsto f(x) = \left\{ \begin{array}{cc}
    f_0(x) = x/(1-x)& x\in \pS_0= [0,1/2]\\
        f_1(x) = (1-x)/x & x \in \pS_1= (1/2, 1]
             \end{array} \right. \, ,
\ee{farey-map}
which has been studied intensively in connection with continued
fraction expansions and
circle maps\cite{maye,MaRo,ACK,PC87,C92a,AACII,Mayer91,sum_rules}.

\BFIG{.9}{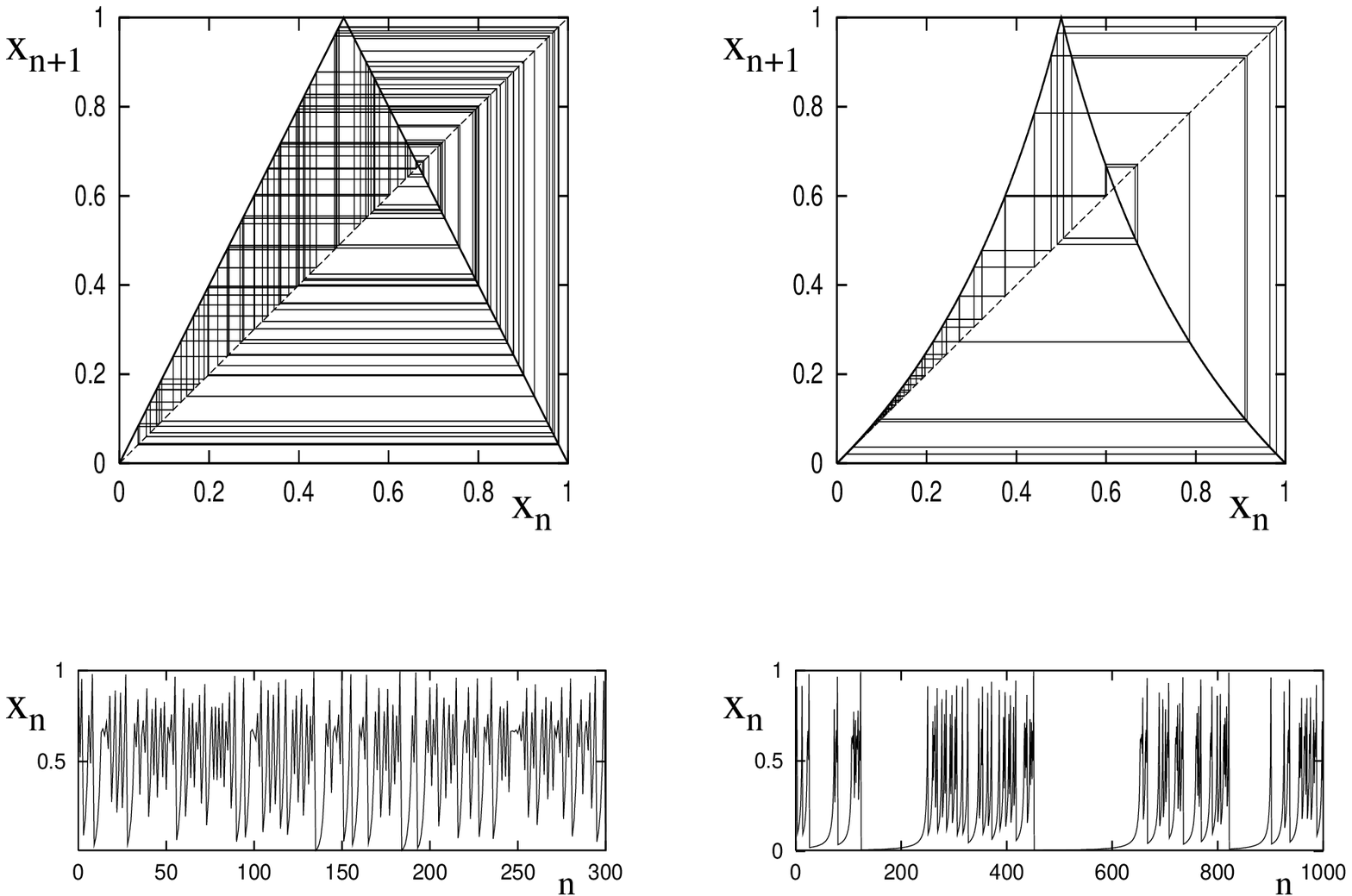}
{}{
(a) A tent map trajectory.
(b) A  Farey map trajectory.
}{tent-farey}
%
The dynamics of the Farey map, which is of the form
\refeq{intereq:mapprot} with $s = 1$, is compared to that of the
hyperbolic, piecewise linear tent map
\beq
x \mapsto f(x) = \left\{ \begin{array}{cc}
    f_0(x) = 2 x& x\in \pS_0=[0,1/2]\\
        f_1(x) = 2 (1-x) & x \in \pS_1= (1/2, 1]
                 \end{array} \right. \,
\ee{tent}
in \reffig{tent-farey}.
In sharp contrast to the uniformly chaotic trajectory of the tent map,
the Farey map trajectory alternates intermittently
between slow regular motion
close to the marginally stable fixed point, and chaotic bursts.
The presence of marginal stability has striking dynamical consequences:
correlation decay may exhibit long range power law asymptotic
behavior and diffusion processes can assume anomalous character whereas
escape from a repeller may be algebraic rather than exponential.
In quantum mechanical systems linked to a classical deterministic
dynamics via periodic orbit trace formula, intermittency can be shown
to lead to localisation and quasi-regular
eigenfunctions\cite{Tan95,Tan96,Tan97}.

The
Ruelle thermodynamical formalism\cite{ruelle-b,PG97,DasBuch}
relates dynamical averages to the
spectral properties of the \FPoper\ or generalizations
thereof, the Ruelle-Araki operators. For a 1-dimensional
map a \FPoper\ relates the density at current time to the density
at the previous time step:
\beq
\rho_{n+1}(x) = \Lop[\rho_n](x) = \int dy\,  \delta(x - f(y)) \rho_n(y)
          = \sum_{f(y) = x} |f'(y)|^{-1} \rho_n(y).
\ee{PF-oper}

Here we shall show how the method of cycle expansions may be
applied to the calculation of spectra
of \FPoper s in the presence of marginally
stable orbits.
We start in \refsect{sec:PF-hyp} by recapitulating the theory for
hyperbolic {maps}, and introduce
\dzeta s, \fd s, and their cycle expansions.
In \refsect{sec:PF-int} we describe
\FPoper s in the presence of isolated marginally stable fixed points and
introduce the induced map of an intermittent map.
The core of the paper is \refsect{sec:cyc-exp-int};
we explain how  marginally stable orbits are incorporated in cycle expansions.
In order to gain some appreciation for the difficulties lying ahead,
we start in \refsect{sec:toy} by studying a piecewise linear map
which follows the scaling \refeq{intereq:mapprot} in the neighborhood of a
fixed point,
show that the \dzeta\ has a branch cut, and extract from this cut the
power law  that characterizes the escape from the repeller.

Such branch cuts are typical also for smooth intermittent
maps with isolated marginally stable fixed points and cycles.
In \refsect{ss-inter-gen} we discuss
higher order curvature corrections required by
their cycle expansions.
The knowledge of the type of singularity one encounters
for such systems enables us
to construct the resummation method presented
in \refsect{ss-inter-resum}. Analytic
continuations of \dzeta s in terms of their integral representations are
discussed in \refsect{ss-inter-integ}.

\section{\FPoper s and cycle expansions}
\label{sec:PF-general}

Even though the operator formalism is not the focus of this paper,
it is instructive to review it here as it sheds light on the cycle
expansion approach pursued later. Briefly, when a dynamical system
is uniformly hyperbolic and its symbolic dynamics is a subshift of finite
type (a complete Smale horseshoe), \Fd s of the associated \evOper s
(like the \FPoper\ \refeq{PF-oper}) are entire and their spectra are
discrete\cite{Rugh92}, very much like the bound spectra of quantum,
electrodynamical and elastodynamic systems. However, if either the
requirement of hyperbolicity, or of the finite subshift, or both are
not fulfilled, the {\Fd} is no longer an entire
function, with delicate consequences for the spectra of
evolution operators\cite{DasBuch,CRR93}.

We are far from having a global theory for non-hyperbolic dynamics; partial
answers can, however, be given relating isolated features in the dynamics to
spectral properties of the \FPoper .
We will focus on a special case of non-hyperbolicity in what follows,
that is, intermittent maps with complete symbolic dynamics and a single
marginal fixed point.
For intermittent dynamics where the symbolic dynamics is either
not known, or cannot be finitely specified, a good practical method is
the {\em stability ordered} cycle expansions\cite{DC97}.
Alternatively, one can make use of statistical methods developed
by Dahlqvist\cite{Dah94,Dah95} in which \Fd s are approximately
described in terms of probability distributions for the return time to
the non-hyperbolic region.

Another example of non-hyperbolicity in the finite subshift setting
is the Ulam map or logistic map $x \to 4x(1-x)$, with strongly contracting
non-hyperbolic critical point $x_c=1/2$.
The Ulam map is conjugate to the tent map and this
conjugacy fails only at the leftmost fixed point, the forward image of the
critical point. It was shown in ref.~\citen{AACII} that for maps, such as
the ``skew'' Ulam map which cannot be conjugated to a tent map,
the spectrum can be obtained by dividing out from the \Fd\ the
fixed point that the critical point iterates to.
 
A similar technique needs to be employed when dealing with intermittency.
As will be discussed later, marginally stable cycles are again factored
out from the \Fd\ leading
in this case, however, to a whole sub-determinant related to the
dynamics close to the marginally stable point. The corresponding operator is
(as can be shown by explicit conjugation) equivalent to free translation
with a continuous spectrum, just
as it is in quantum scattering where asymptotic states are also translation
invariant. What remains of the factored \Fd\ can be interpreted as a \Fd\ of
another dynamical system, the \Fd\ of a purely hyperbolic
{\em induced map}.

\subsection{The hyperbolic case - a review}
\label{sec:PF-hyp}

In what follows we shall illustrate the key techniques
by considering the dynamics of 1-dimensional hyperbolic maps
$x \mapsto f(x)$
on the unit interval
 $\pS = [0,1]$
with two piecewise-analytic monotone branches
\begin{equation}
f(x)=\left\{ \begin{array}{ll}
f_0(x) &  x \in \pS_0= [0,a] \\
f_1(x) &  x \in \pS_1=\, (b,1] \end{array}
     \right.
\,.
\label{int.pl}
\end{equation}
The two branches are assumed complete, $f_0(\pS_0)=f_1(\pS_1)=\pS$.
Such a map allows for escape if $a<b$, and the dynamics
is bounded to the unit interval if $a=b$. Its
symbolic dynamics is the full binary shift, with a
1-to-1 mapping between admissible orbits and
all possible infinite binary itineraries.
A map is hyperbolic if the stability exponents of all periodic orbits
are strictly positive
\beq
\frac{1}{n_p} \ln |\Lambda_p| =
    \frac{1}{n_p} \, \sum_{i=1}^{n_p}\ln |f'(x_i)| \ge c > 0
\,,
\ee{cycle-weights}
where the sum is over the cycle points $x_i$ along the cycle $p$.
In this case a well established theory exists\cite{ruelle-b,Rugh92},
and it can be shown that the \FPoper s have discrete spectra
$ \{\lambda_0, \lambda_1, \lambda_2 \ldots \}$ when acting on
a suitable space of real analytic functions.
The method of {\em cycle expansions}\cite{PC89,AACI,DasBuch}
is often an efficient method of computing the spectrum.
It makes use of
the fact that the trace of the \FPoper\ can be written in terms of
the periodic orbits of the map,
\beq
\tr\Lop^n = \sum_{\alpha} \lambda_{\alpha}^n =
\int_0^1 dx \,\delta(x - f^n(x)) =
\sum_i^{(n)} \frac{1}{|1 - \Lambda_i|} \, ,
\ee{trace}
where the sum is taken over all periodic points of period $n$.
The \fd\ of $\Lop$ is related to the traces by the
cumulant expansion\cite{DasBuch},
\bea \label{zeta-def}
\det \left(1 - z \Lop\right)
     &=& 1 - z \, \tr \Lop -
     \frac{1}{2} z^2 \left[\tr \Lop^2-(\tr \Lop)^2\right]-\ldots
    \continue
     &=& \prod_{k=0}^{\infty}
     \prod_p \left(1 - \frac{z^{n_p}}{|\Lambda_p| \Lambda^k_p}\right)
     = \prod_{k=0}^{\infty} \zfct{k}(z)
\,,
\eea
where the product runs over all prime periodic orbits
(repetitions excluded) of periods $n_p$.
For hyperbolic maps with complete binary symbolic dynamics,
the \fd\ is an entire function\cite{ruelle-b,Rugh92}
whose zeros $\{z_0, z_1, z_2,
\ldots \}$ are related to the eigenvalues of the \FPoper\ via
$\lambda_{\alpha} = z^{-1}_{\alpha}$.

A quantity easily accessible to experiment and numerical calculations
is the survival probability $\hat{\Gamma}_n$,  that is, the fraction
of phase space which remains within
the phase space volume $\pS$ (with $\pS = [0,1]$
for the map (\ref{int.pl})) for at least $n$ iterations
\beq
\hat{\Gamma}_n = {1 \over |\pS|} \sum_{i}^{(\cl{})}\, |\pS_i|
\,.
\ee{escape-rate}
Each surviving interval $\pS_i$ contains
a periodic point $x_i \in \pS_i$ belonging
to a prime cycle $p$ or its $r$th repeat, of total
period $r\cl{p} = n$.
For hyperbolic maps their size can be asymptotically
bounded from below and
above in terms of the stabilities of that orbit,
\beq
{\cal C}_1 \frac{1}{|\Lambda_p|}
< \frac{|\pS_i|}{|\pS|} <{\cal C}_2 \frac{1}{|\Lambda_p|}
\,,
\ee{bounds-hyp}
with ${\cal C}_1, {\cal C}_2$ independent of $p$. The asymptotic
behavior of $\hat{\Gamma}_n$ is thus captured by
the periodic orbit sum
\beq
\hat{\Gamma}_n \sim \Gamma_n = \sum_{i}^{(n)}\, \frac{1}{|\Lambda_i|}
\, .
\ee{escape-hyp}
For strictly hyperbolic maps
this survival probability
falls off exponentially with $n$
in the large $n$ limit.
The periodic orbit sum (\ref{escape-hyp}) can be rewritten
in terms of $\zetaInv_{0}$ defined in (\ref{zeta-def}),
\begin{equation}
\Gamma_n=\frac{1}{2\pi i} \oint_{\gamma_r^-} z^{-n}
\left( \frac{d}{dz} \log \zeta_0^{-1}(z) \right) dz \, ,
\label{intereq:repr1}
\end{equation}
where the contour encircles the origin in the clockwise
direction and lies inside the unit circle $|z| = 1$. The
\dzeta\ $\zetaInv_0$ can be written as a convergent product
over periodic orbits in this region, integrals and sums can be interchanged,
and the formula (\ref{escape-hyp}) is recovered after
evaluating the integrals term by term by Cauchy integration.
For hyperbolic maps, cycle expansions
(discussed in more detail below) provide an analytic extension
of the \dzeta\ beyond the leading zero; we may then
deform the original contour into a larger circle with radius $R$ which
encircles both poles and zeros of $\zetaInv_0$.
Residue calculus turns this into a sum over the zeros $z_{\alpha}$ and poles
$z_{\beta}$ of the \dzeta,
\begin{equation}
\Gamma_n=
\sum^{\mbox{\footnotesize zeros}}_{|z_\alpha|<R}
{1 \over z_\alpha^{n}}
-
\sum^{\mbox{\footnotesize poles}}_{|z_\beta|<R}
{1 \over z_\beta^{n}}
+\frac{1}{2\pi i} \oint_{\gamma_R^-} dz\, z^{-n}
\frac{d}{dz} \log \zeta_0^{-1} ,
\label{intereq:repr2}
\end{equation}
where the last term gives a contribution from a large
circle $\gamma_R^-$. We thus find exponential decay of $\Gamma_n$
dominated in general by the leading zero $z_0$ of
$\zetaInv_0(z)$. We expect $z_0= 1$ for
bounded dynamics and $|z_0| > 1$ for maps which allow for
escape.

The all important caveat here is whether one is able
to construct an analytic continuation of $\zetaInv_0(z)$ for
$|z| > |z_0|$.
For systems with a complete symbolic dynamics with
finite alphabet the method of cycle expansions\cite{AACII},
with cycles and products of cycles
(pseudo-cycles) ordered by their total period,
has been demonstrated to work very well.
For maps
with a binary symbolic dynamics one has
\bea \label{cyc-exp-bin}
 \zetaInv_0(z) &=& \prod_p \left(1 - \frac{z^{n_p}}{|\Lambda_p|}\right) =
                     1 - \sum_{n=1}^{\infty} c_n\\ \nonumber
 &=& 1 - z \left[\frac{1}{|\Lambda_0|} + \frac{1}{|\Lambda_1|}\right]
       - z^2 \left[\frac{1}{|\Lambda_{01}|} -
       \frac{1}{|\Lambda_0 \Lambda_1|} \right]
       - z^3 \left[\cdots \right]\,-\, \cdots
\eea
Here the $c_1$ term,
the {\em fundamental} term,
includes all unbalanced, not shadowed cycles.
For the higher terms,
called {\em curvature} corrections, large cancellations
between cycles and pseudo-cycles that shadow them take
place\cite{AACII,DasBuch}.

For a piecewise linear map like the tent map \refeq{tent} with
$|\Lambda_p| = 2^{n_p}$ for all periodic orbits of length $n_p$,
contributions from cycles and pseudo-cycles
cancel exactly and the curvature contributions are identically zero.
The \dzeta\ thus has the form
\beq
 \zetaInv_0(z) = 1 - z
\ee{tentZeta}
and vanishes  at $z_0 = 1$, as expected. In this case
the complete spectrum of the \FPoper\
$\{ \lambda_0, \lambda_1, \lambda_2, \ldots\}$
can easily be deduced from a cycle expansion of higher order
\dzeta s  $\zetaInv_k(z)$; one obtains
$\lambda_\alpha = z_\alpha^{-1} = 1/2^{2 \alpha}$,
$\alpha = 0,1,2, \ldots$.

\subsection{The intermittent case and the induced map}
\label{sec:PF-int}

For intermittent maps of the form \refeq{intereq:mapprot}
the trace and determinant
formulas (\ref{trace}), (\ref{zeta-def}) have no meaning as they stand;
the periodic orbit stabilities can be arbitrarily close or equal to 1,
and that renders the formulas singular.
A spectral analysis of the \FPoper\ is now much subtler than in the
hyperbolic case.
In this section we briefly sketch the main ideas behind the
recent breakthroughs in the spectral analysis for
intermittent maps\cite{intRugh,intIso,SIF,TPF}
which establish that the
\FPoper\ has a continuous spectrum $\sigma_c = [0,1]$, and a pure point
spectrum
consisting of isolated eigenvalues of finite multiplicity.
Due to the symmetry of the map in this case, zero is an
eigenvalue of infinite multiplicity.
We shall return to \dzeta s and \fd s in \refsect{int-branch} and will
show there that marginal orbits induce branch cuts in zeta
functions which are related to the continuous part of the spectrum.

The starting point for a rigorous analysis is the precise specification of
the function space that the \FPoper\ acts on.
This requires a level of analysis that lies beyond the scope of
this paper: in particular, one needs to prove that $\Lop$ maps
this function space into itself. In what follows we will only give
the key steps of the operator approach, which will be instructive later
when discussing cycle expansions, without attempting to give details of
the proofs.

The basic idea is to separate the intermittent part of the dynamics
from the chaotic dynamics. For clarity, we
shall illustrate the method using the Farey map
(\ref{farey-map}) as an example\cite{SIF,TPF}.
Denote by $F_i(x), \,\, i\in \left\{0,1\right\}$ the inverse of
the branch $f_i$ with $F_0(x) = x/(x+1)$ and
$F_1(x) = 1/(x+1)$. $F_i$ maps the unit interval $\pS$ into
$\pS_i$, the interval supporting $f_i$.
For the Farey map, the \FPoper\ \refeq{PF-oper} acts on the
two branches as
\bea
&&\Lop = \Lop_0 + \Lop_1
\,, \continue
&& \Lop_0[\rho](x) = \frac{1}{(1+x)^2}\, \rho\left(\frac{x}{x+1}\right)
\,,\quad
\Lop_1[\rho](x) = \frac{1}{(1+x)^2}\, \rho\left(\frac{1}{x+1}\right)
\,.
\label{PF-operFarey}
\eea
We focus on the intermittent behavior of the
$0$ - branch by formally factoring it out
\beq
\left(1 -z \Lop \right) =
  (1 - \hat{\Lop}(z)) \cdot  \left( 1 -z \Lop_0 \right)
\,.
\ee{splitL}
Expanding
\[
(1 - \hat{\Lop}(z)) = (1-z\Lop_0 -z \Lop_1){ 1 \over  1 -z \Lop_0 } =
    1 - z \Lop_1{ 1 \over  1 -z \Lop_0 }
\]
as a geometric series yields a formula for $ \hat{\Lop}$:
\beq
  \hat{\Lop}(z) = \sum_{m=1}^{\infty} \, z^{m} \Lop_1 \Lop_0^{m-1}
\,.
\ee{Mz}
Loosely speaking,
the partial \FPoper\ $\Lop_0$
in the decomposition \refeq{splitL}
is responsible for the continuous part of the spectrum,
a new feature specific to intermittent dynamics, while
$\hat{\Lop}$ adds isolated eigenvalues.

The operator $\Lop_0$ is conjugated to the shift operator
$S[\varphi](x) = \varphi(1+x)$ under the
transformation $\rho(x) = {\cal C}[\varphi](x) = \varphi(f_1(x))$. The
shift operator has a continuous spectrum $\sigma_c = [0,1]$
on a function space of analytic functions obtained by a generalized
Laplace transform
\[
\varphi(x) = {\cal F}[\psi](x) = \frac{1}{x^2}\,\int_0^{\infty}\,
d\mu(s)\,e^{-s/x}e^{-s}\psi(s)
\]
acting on the space of square integrable functions
$\psi \in L^2(\reals_{+},\mu)$ with $d\mu(s) = s/(e^s - 1) \, ds$ \cite{SIF}.
Similar relations are expected to hold in neighborhoods of marginally stable
fixed points  for general intermittent maps\cite{intRugh}.

It remains to understand the role of  $\hat{\Lop}$.
The meaning of the formal power series
(\ref{Mz}) is best grasped by setting $z=1$: the operator
\[
\hat{\Lop}(1) \, = \,
\hat{\Lop}    \, = \,
    \sum_{m=1}^{\infty}\, \Lop_1 \Lop_0^{m-1}
\]
is the \FPoper\ for the
map\cite{AACII}
\beq
\hat{f}(x)=\left\{ \begin{array}{ll}
f_1(x) &  x \in  \pS_1\\
f_2(x) = f_1(f_0(x)) & x \in \pS_2 = (q_2,q_1]\\
\vdots & \vdots\\
f_m(x) = f_1(f^{(m-1)}_0(x)) & x \in \pS_m = (q_m,q_{m-1}]\\
\vdots & \vdots
\end{array}
\right.  \,,
\ee{ind-map}
constructed from compositions of the initial 2-branch map $f$
defined in \refeq{int.pl}.
The interval boundaries $q_n$ are successive
preimages
$f_0(q_n) = q_{n-1}$ of the rightmost point of $\pS_0$, that is,
$q_1 = 1/2$ for the Farey map or $q_1 = a$ in case
of the map \refeq{int.pl}.
The map
$\hat{f}$ thus has an infinity of branches,
taking care of all sequences of
consecutive iterations of the left branch
of $f$ in a single step. The length of a cycle
of $\hat{f}$ thus equals the number of $1$'s in the
binary itinerary of the original map $f$.
For the Farey map the map
(\ref{ind-map})
is the Gauss map of number theory\cite{Mayer91}
\[
\hat{f}_{Gauss}(x) =
\left\{ \frac{1}{x} \right\} \qquad  x \in (0,1],
\]
where $\{\cdots\}$ denotes the fractional part.
This map is related by conjugacy to
the {\em induced map} of
ref.~\citen{intPrell,PS92}, the first return
map on interval $[1/2,1]$.
The key issue is that the induced map is hyperbolic, so the
operator $\hat{\Lop}$ adds only isolated eigenvalues
to the spectrum.

Now, factoring out parts of the \fd\ implicit in \refeq{splitL} does
not necessarily make sense. If the \fd\ is an entire function, as it
is for hyperbolic maps, this merely introduces singularities in the
remainder term and the spectrum of $\Lop_0$ may have little in common with
the spectrum of the full \FPoper\ $\Lop$. Things are, however, different
for intermittent maps. The determinant $\det(1 - \hat{\Lop}(z))$ is
analytic in ${\complex} - [1, \infty)$, with a branch cut along
the real line $[1,\infty)$, and can be analytically continued through
this branch cut, albeit onto different Riemann sheets. Hence
$\sigma_c = [0,1]$ is part of the spectrum of the full operator
$\Lop$, while the rest of the spectrum of $\Lop$ is discrete and
can be deduced from $\hat{\Lop}$. From (\ref{splitL}) it follows now that
$z^{-1}$ is an eigenvalue of $\Lop$
if $1$ is an eigenvalue of $\hat{\Lop}(z)$.

Let us finally return to the \Fd\ of $\hat{\Lop}$.
The dynamics of $f$ and $\hat{f}$ are equivalent apart from
singling out the fixed point at $x=0$.
As a consequence, very little changes when writing the \fd\
\refeq{zeta-def} for $\hat{\Lop}(z)$ in terms of periodic orbits
except for taking out the marginal fixed point, that is,
\beq
\det \left(1 - \hat{\Lop}(z)\right) = \prod_{k=0}^{\infty}
    \prod_{p\neq 0} \left(1 - \frac{z^{n_p}}{|\Lambda_p| \Lambda^k_p}\right)\, ,
\ee{zetaind-def}
where $n_p$ is again the length of the periodic orbit with respect
to the original map $f$. That something more profound is happening
becomes clear when we consider the first few terms in the cumulant expansion
of the \fd\, Eqn.\ \refeq{zeta-def},
\beq
\det \left(1 - \hat{\Lop}(z)\right) = 1 -\tr \hat{\Lop}(z) - \ldots
= 1 - \sum_{m=1}^{\infty} \frac{z^m}{|1 - \Lambda_{10^{m-1}}|} - \ldots
\, .
\ee{cum-int}
The infinite sum over orbits $10^{m-1}$ reflects the fact that the
induced map has infinitely many fixed points.

\section{Periodic orbit expansions for intermittent maps}
\label{sec:cyc-exp-int}

In order to get to know the kind of problems which arise when studying
\dzeta s in the presence of marginal stability
we shall train on a carefully crafted piecewise linear map first,
and return to the case of smooth intermittent
maps in \refsect{ss-inter-gen}.

\subsection{A piecewise linear toy map}
\label{sec:toy}

The tent map (\ref{tent}) is an idealized example of
a hyperbolic map.  To construct a piecewise linear
analogue for intermittency,
we start from the induced map
(\ref{ind-map}), linear in each branch\cite{intGasWan}.
\SFIG{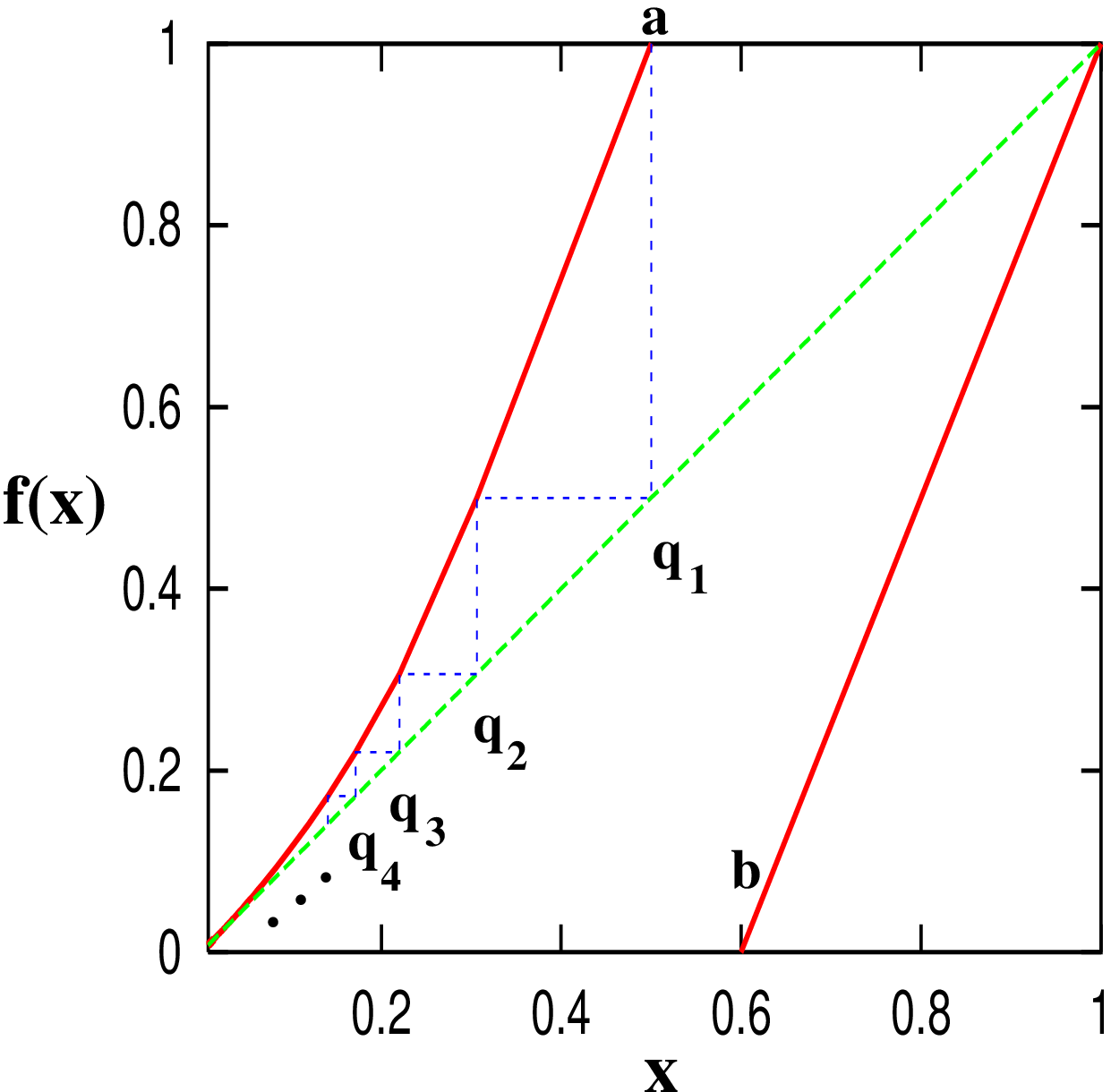}
{}{
A piecewise linear intermittent map of type (\ref{int.pl})
for intervals chosen according to \refeq{in-def} with
$a = .5$, $b = .6$, $s=1.0$.
}{int:plm}
%

We take the right branch in (\ref{int.pl}) expanding and linear:
\[
f_1(x)=\frac{1}{1-b} (x-b) \quad \mbox{for} \quad x \in \pS_1 = (b,1]
\,.
\]
The left branch $f_0$ of $f$ is constructed in such a way
that it will exhibit
the intermittent behavior (\ref{intereq:mapprot}) near
the origin. Motivated by the form of the induced map
(\ref{ind-map}),
we consider a monotonically decreasing sequence $q_m$ of
preimages of the rightmost point of $f_0$ in
$\pS_0 = [0,a]$, with $q_1=a$ and $q_m \rightarrow 0 $ as
$ m \rightarrow \infty $. This sequence defines a partition of the
left interval $\pS_0$ into an infinite number
of monotonically decreasing intervals $\pS_m$, $m\ge 2$ with
\begin{equation}
\pS_m=  \, (q_{m}, q_{m-1}] \qquad \mbox{and} \qquad
\pS_0 = \bigcup_{m=2}^{\infty} \pS_m  .
\label{intereq:indef}
\end{equation}

Conversely, any given monotone sequence
$\{ q_m \}$ {\em defines} a piecewise linear continuous
map $f_0(x)$, with
the slopes of the various linear segments given by
\beq
\begin{array}{rccl}
f_1'(x)&=&\frac{1}{1-b}
          & \quad \mbox{for} \,\, x \in \pS_1
        \\
f_0'(x)
    &=&
      \frac{1-a}{|\pS_2|}
          & \quad \mbox{for} \,\, x \in \pS_2
        \\
f_0'(x)&=&
      \frac{|\pS_{m-1}|}{|\pS_{m}|}
          & \quad \mbox{for} \,\, x \in \pS_{m}, \,\, m\geq 3
\end{array}
\ee{lisDahlEqs}
with $|\pS_{m}| = q_{m-1} - q_{m}$ for $m > 1$.

We have seen in \refeq{ind-map} and \refeq{cum-int} that the family of
periodic orbits with
itinerary $1 0^{m-1}$ plays a key role for intermittent
maps of the form (\ref{intereq:mapprot}).
An orbit $1 0^{m-1}$ enters the intervals
$\pS_1 \to \pS_{m} \to \pS_{m-1}, \ldots \to \pS_2$ successively and the
family comes closer and closer to the marginally
stable fixed point at $x = 0$ as $m \to \infty$.
The stability of a cycle $1 0^{m-1}$ for $m \ge 2$ is
proportional to the size of the interval closest to the
marginal fixed point
\beq
\Lambda_{10^{m-1}} = f '_0(x_{m})  f '_0(x_{m-1}) \ldots
f '_0(x_2)  f '_1(x_1)
        =  \frac{1}{|\pS_{m}|}\frac{1-a}{1-b} \,,
\ee{stab10n}
with $x_i \in \pS_i$.

A piecewise linear map of the form (\ref{int.pl})
is fixed by
the sequence $\{q_m\}$. By choosing $q_m = 2^{-m}$, for example, we
recover the uniformly hyperbolic Bernoulli shift map.
An algebraically decaying sequence
$\{ q_m\}$  
\beq
q_m\sim \frac{1}{ m^{1/s}}
\,,\qquad
|\pS_m| \sim\frac{1}{m^{1+1/s}}
\,,
\ee{in-asym}
yields an intermittent map  of the form (\ref{intereq:indef}),
with the asymptotic behavior controlled by
the intermittency exponent $s$ in (\ref{intereq:mapprot}).
The stability eigenvalues of periodic orbit families
approaching the marginally stable fixed point,
such as the $10^{m-1}$ family in \refeq{stab10n}, grow
in turn only algebraically (not exponentially) with the cycle length.

It may now seem natural to construct an intermittent toy map in terms of
a partition $|\pS_m| = 1/m^{1+1/s}$, that is, a partition
which follows \refeq{in-asym} exactly. Such a choice leads to a
\dzeta\  which can be written in terms of
Jonqui\`ere functions (or polylogarithms)\cite{FK,Htf1}.
We shall, however, use the freedom in
choosing the partition to make life as simple as possible
later on, without loosing the key intermittency features.
Inspired by a bit of reverse engineering we fix the intermittent
map intervals $\pS_m$ by
\beq
|\pS_m|={\cal C}\, \frac{\Gamma(m+\ell -1/s-1)}{\Gamma(m+\ell)}
\qquad \mbox{for} \quad m\ge 2,
\ee{in-def}
where $\Gamma(x)$ is the Gamma function, $\ell = [1/s]$
denotes the integer part of $1/s$,
and the normalization constant $\cal C$ is
fixed by the full partition condition
$\sum_{m=2}^{\infty} |\pS_{m}| = q_1 = |\pS_0|$, that is,
\beq {\cal C} =
|\pS_0| \left[\sum_{m=l+1}^{\infty}
\frac{\Gamma(m-1/s)}{\Gamma(m+1)}\right]^{-1}.
\ee{norm-inter}
One easily verifies that the intervals decay asymptotically like $m^{-(1+1/s)}$
as required by the condition (\ref{in-asym})
using Stirling's formula for the Gamma function.

The point spectrum of $f$ is given by the zeros of the \fd\
(\ref{Mz}) or the associated \dzeta s.
In order to find analytic continuations of the \dzeta\ $\zetaInv_{k}$,
we start by expanding the product as we did in (\ref{cyc-exp-bin}).

Due to the absence of the
fixed point $0$, there are no
pseudo-cycles that shadow cycles of the form $10^{m-1}$,
so the fundamental cycles are now the entire infinite family of
cycles with itineraries of the form $10^{m-1}$ accumulating towards
the marginal fixed point $\overline{0}$. As the $\overline{0}$
fixed point does not participate in the dynamics, we have to
switch from the 2-letter $\{0,1\}$ alphabet to the infinite alphabet
$
m = 10^{m-1}
$ which labels the branches of the induced map
(\ref{ind-map}).
The admissible itineraries are now generated by all walks on the
infinite Markov diagram\cite{AACII} \reffig{finter_mark}.
%
\SFIG{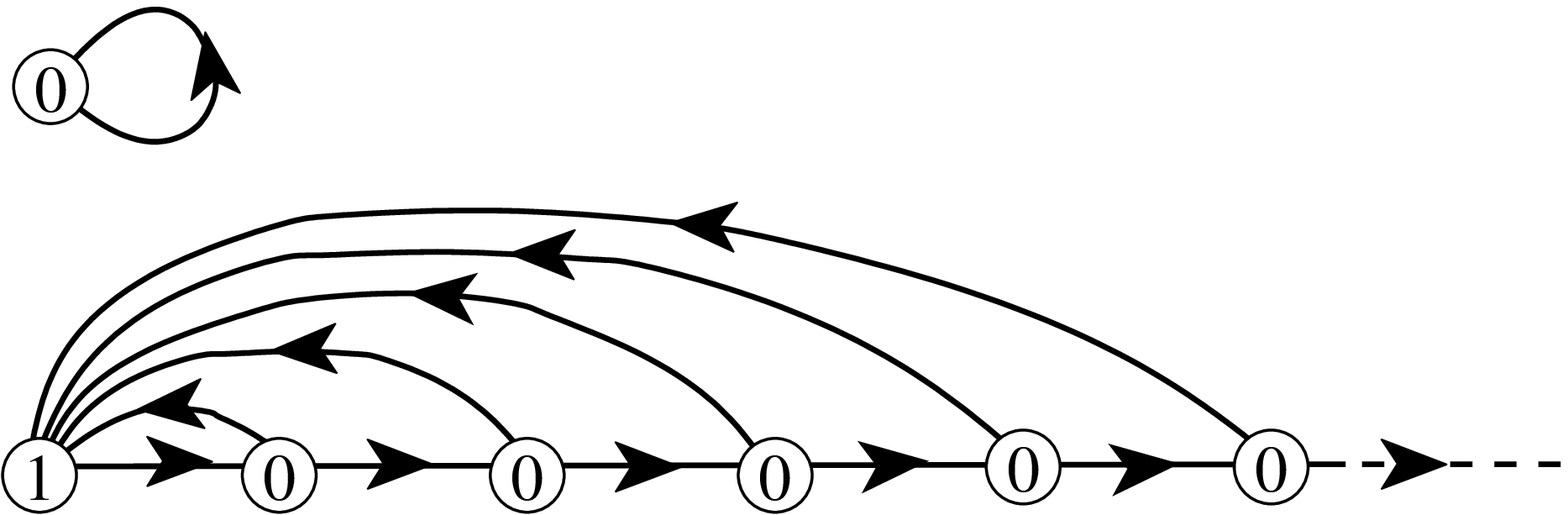} 
{}{
\MarkGraph\ for the infinite alphabet
$ \{m= 10^{m-1} ; \overline{0}\,,\,\, m\geq 1\}$
derived from full binary symbolic dynamics by separating out
the $\overline{0}$ fixed point.
}{finter_mark}
%

The piecewise linear form of the map which maps intervals $\pS_m$
exactly onto $\pS_{m-1}$ simplifies matters considerably: when
expanding the zeta functions in the spectral determinant \refeq{zetaind-def}
all cycles which traverse the right branch at
least twice are cancelled exactly by pseudo-cycles. The cycle expanded
\dzeta s depend thus only on the fundamental cycles and have the form
\beq
\zetaInv_{k}(z) =
\prod_{p\neq 0} \left( 1-\frac{z^{n_p}}{|\Lambda_p|\Lambda_p^k} \right)
=1-\sum_{m=1}^{\infty}
\frac{z^{m}}{|\Lambda_{10^{m-1}}|\Lambda_{10^{m-1}}^k} \, .
\ee{cyc-exp-piecelin}
To keep the discussion simple, we will restrict our considerations
to $\zetaInv_{0}$.
We obtain
\beq
\zetaInv_{0}(z) = 1-  (1-b) z- {\cal C}\frac{ 1-b}{1- a}
    \sum_{m=2}^{\infty} \frac{\Gamma(m+\ell-1/s -1)}{\Gamma(m+\ell)} z^m
\,.
\ee{cyc-exp-toy}
The cycle expansion does not, however, lead to an analytic
continuation here, despite the exact cancellation between
cycles and pseudo-cycles, in contrast
to the \dzeta\ for the tent map \refeq{tent}.
Due to the slow, algebraic decay of cycle weights,
the sum (\ref{cyc-exp-toy}) is divergent for $|z| > 1$.
We shall now show
that this behavior is due to a branch cut in $\zetaInv_{k}(z)$
along $[1,\infty)$.

\subsection{Branch cuts in \dzeta s due to intermittency}
\label{int-branch}
In order to analytically continue the sum (\ref{cyc-exp-toy})
beyond $|z| \ge 1$
we make use of the following
binomial identities valid for $ \alpha = 1/s >0$:
\bea
\alpha \mbox{ non-integer} &\quad&
(1 - z)^{\alpha} =
\sum_{n=0}^{\infty}
\frac{\Gamma(n-\alpha)}{\Gamma(-\alpha)\Gamma(n+1)} z^n
                        \label{bra-int}\\
\alpha \mbox{ integer} &\quad&
    \begin{array}{rcl}
(1 - z)^{\alpha}\log(1-z) &=&\sum_{n=1}^{\alpha} (-1)^n c_n z^n
\\ 
&&+ (-1)^{\alpha+1} \alpha! \sum_{n=\alpha+1}^{\infty}
\frac{(n-\alpha-1)!}{n!} z^n
    \end{array}
\continue
 \mbox{ with}&\quad&
c_n =  {\alpha \choose n} \sum_{k=0}^{n-1} \frac{1}{\alpha-k}\, .
\nnu
\eea
For simplicity sake, we restrict
the intermittency parameter to
$1\le 1/s<2$, that is the case
$[1/s] = \ell = 1$. All that follows can easily
be generalized to arbitrary $s > 0$ using
\refeq{bra-int}.
What once was a mystifying choice of interval widths (\ref{in-def})
now pays off:
the infinite sum (\ref{cyc-exp-toy}) can be evaluated analytically,
\[
\sum_{m=2}^{\infty} \frac{\Gamma(m-1/s)}{\Gamma(m+1)} z^m = \left\{
\begin{array}{lcl}
\Gamma(-\frac{1}{s})\left[(1-z)^{1/s} - 1 + \frac{1}{s} z\right]&
\mbox{for}&1<1/s <2;\\
(1-z)\log(1-z) + z & \mbox{for}&s = 1 \; .
\end{array} \right.
\]
The normalization constant $\cal C$  in (\ref{norm-inter}) can be
evaluated, and  one obtains for
$1<1/s <2$:
\beq
\zetaInv_{0}(z) =
    1-(1-b)z- \frac{a}{1/s-1} \frac{1-b}{1-a}
        \left( (1-z)^{1/s} - 1 + \frac{1}{s} z\right)
\ee{zeta-toy-ana1}
and for $s = 1$:
\beq
\zetaInv_{0}(z) =
    1-(1-b)z- a \frac{1-b}{1-a}
                \left((1-z)\log(1-z) + z \right) \, .
\ee{zeta-toy-ana2}
For general $s >0$ one obtains
\beq
\zetaInv_{0}(z) =
    1-(1-b)z- \frac{a}{g_s(1)} \frac{1-b}{1-a}
    \frac{1}{z^{\ell-1}}\left(
(1-z)^{1/s} - g_s(z)\right)
\ee{zeta-toy-ana-gen1}
for non-integer $s$ with $\ell = [1/s]$ and
\beq
\zetaInv_{0}(z) =
    1-(1-b)z- \frac{a}{g_s(1)}
\frac{1-b}{1-a}\frac{1}{z^{\ell-1}}\left((1-z)^{\ell}\log(1-z)- g_s(z) \right)
\ee{zeta-toy-ana-gen2}
for $1/s = \ell$ integer; here, $g_s(z)$ are polynomials of order $\ell=[1/s]$
which can be deduced from \refeq{bra-int}.
We see that the \dzeta\ has a branch cut
starting at $z =1$ and running along the positive real
axis. We find algebraic branch cuts for non integer intermittency
exponents $1/s$ and logarithmic branch cuts for $1/s$, integer. We will see
in \refsect{ss-inter-gen} that branch cuts of that form are generic for
1-dimensional intermittent maps.
In the next section, we will use these analytic expressions to calculate
the escape from our toy map for $a < b$ in \refeq{int.pl}.

\subsection{Escape rate}
The fraction of survivors $\Gamma_n$ defined in (\ref{escape-rate}) have
for hyperbolic maps been linked to $\zetaInv_{0}$ with the help
of bounds given in (\ref{bounds-hyp}). These bounds are no longer valid
for intermittent maps. It turns out, however, that bounding survival
probabilities strip by strip is a far too strict requirement for
establishing the relation  (\ref{escape-hyp}).
In fact, a somewhat weaker bound can be established,
linking the average size of intervals {\em along a periodic orbit} to
the stability of the periodic orbit for all but the interval $\pS_{0^n}$.
We can write this bound by averaging over each prime cycle $p$ separately,
\beq
{\cal C}_1 \frac{1}{|\Lambda_p|} <
\frac{1}{n_p} \sum_{p} \frac{|\pS^{(n_p)}_{p}|}{|\pS|}
<{\cal C}_2 \frac{1}{|\ExpaEig_p|} ,
\ee{intereq:bound1weak}
for some positive constants ${\cal C}_1$, ${\cal C}_2$
independent of $p$. A proof, which relies on the hyperbolicity
of the induced map, can be found in ref.\ \citen{PDalhlqProof}.
Summing over all periodic orbits leads then again to (\ref{escape-hyp})
which is linked to $\zetaInv_{0}$ via the integral representation
\refeq{intereq:repr1}.

\SFIG{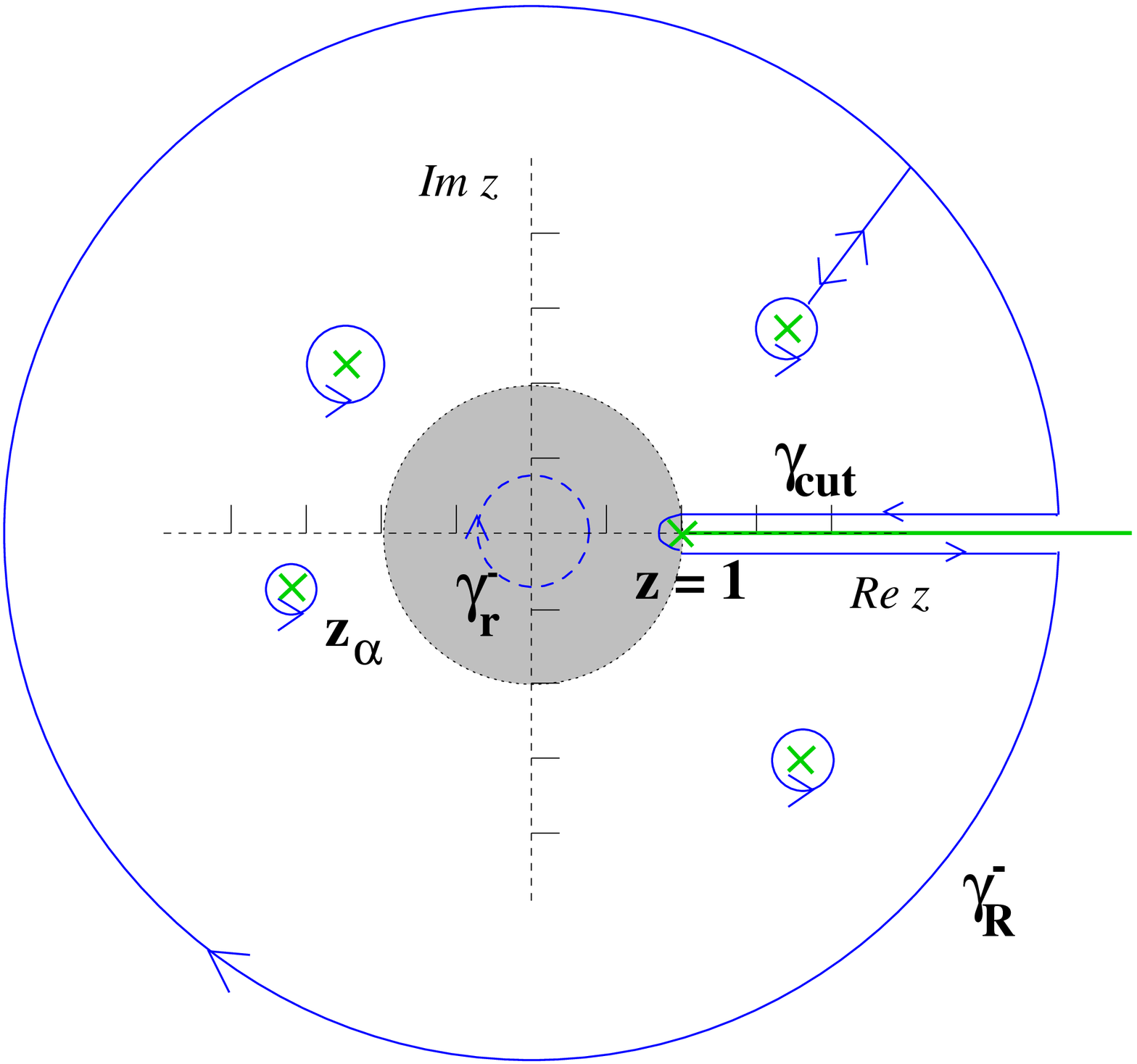}{}
{The survival probability $\Gamma_n$ calculated by contour integration;
integrating (\ref{intereq:repr1}) inside the domain of
convergence $|z| < 1$ (shaded area) of $\zetaInv_0(z)$ in periodic orbit
representation yields
(\ref{escape-hyp}). A deformation of the contour $\gamma_r^-$ (dashed line)
to a larger circle $\gamma_R^-$ (without crossing the branch cut) gives
contributions from the poles and zeros (x) of $\zetaInv_0(z)$ between the
two circles, as well as contributions from the integration along the
branch cut $\gamma_{cut}$.}
{finter_cont1}

The path deformation to $|z| > 1$ that led to (\ref{intereq:repr2})
requires more care here, as it must not cross the branch cut along
$ [1,\infty]$. When
expanding the contour to large $|z|$ values, we have to deform it
in such a way that it sandwiches the branch cut in anti-clockwise direction,
see \reffig{finter_cont1}.  Denoting the detour around the cut as
$\gamma_{cut}$,  we may write symbolically
\[
\oint_{\gamma_r}
=
\sum^{\mbox{\footnotesize zeros}}
-\sum^{\mbox{\footnotesize poles}}
+\oint_{\gamma_R} +\oint_{\gamma_{cut}}
\]
where the sums include only the zeros and the poles in the area enclosed
by the contours.

Let us now go back to our intermittent toy map. The asymptotics of the
survival probability of the map is governed by the behavior of the
integrand $\frac{d}{dz}\log\zeta_0^{-1}$ in (\ref{intereq:repr1}) at the branch
point $z = 1$. We restrict ourselves again to the case $1< 1/s <2$ first
and write the \dzeta\ (\ref{zeta-toy-ana1}) in the form
\[
\zetaInv_0(z)= a_0 + a_1 (1-z) + b_0 (1-z)^{1/s} \equiv G(1-z)
\]
with
\[ a_0 = \frac{b-a}{1-a}, \quad b_0 = \frac{a}{1-1/s} \frac{1-b}{1-a}.\]
Setting $u = 1-z$, we need to evaluate
\beq
\frac{1}{2\pi i}
\oint_{\gamma_{cut}} (1-u)^{-n} 
\frac{d}{d u}
\log G(u)  du   \,
\ee{intereq:intlogG}
where $\gamma_{cut}$ goes around the cut (\ie, the negative $u$ axis).
Expanding the integrand $\frac{d}{d u} \log G(u) = G'(u)/G(u)$ in powers
of $u$ and $ u^{1/s}$ at $u=0$, one obtains
\beq
\frac{d}{d u}
\log G(u)= \frac{a_1}{a_0}+\frac{1}{s}\frac{b_0}{a_0}u^{1/s-1} +O(u)\,   .
\ee{intereq:logG}

The integral along the cut may be evaluated using the general formula
\beq
\frac{1}{2\pi i}\oint_{\gamma_{cut}}
u^\alpha (1-u)^{-n} du=
\frac{\Gamma(n-\alpha-1)}{\Gamma(n)\Gamma(-\alpha)}
\sim  \frac{1}{n^{\alpha+1}} (1+O(1/n))
\ee{intereq:useful1}
which can be obtained by deforming the contour back to a loop around
the point $u =1$, now in positive (anti-clockwise) direction. The
contour integral then picks up the $(n-1)$-st term in the
Taylor expansion of the function $u^{\alpha}$ at $u =1$,
see \refeq{bra-int}.

Inserting  (\ref{intereq:logG}) into (\ref{intereq:intlogG})
and using (\ref{intereq:useful1}) we get the asymptotic result
\beq
\Gamma_n\sim \frac{b_0}{a_0}\frac{1}{s}\frac{1}{\Gamma(1-1/s)}
\frac{1}{n^{1/s}} =
\frac{a}{s-1}\frac{1-b}{b-a} \frac{1}{\Gamma(1-1/s)}
 \frac{1}{n^{1/s}}
\,.
\ee{intereq:interesc}
We see that, asymptotically, the escape from an
intermittent repeller is described by power law decay rather than the
exponential decay we are familiar with for hyperbolic maps. The zeros
and poles of $\zetaInv_{0}$ give exponentially decreasing contributions,
which may indeed be dominant over large time interval before the
power law decay sets in for $n\to\infty$. A numerical simulation of
the power-law escape from an intermittent repeller is shown
in \reffig{finter_cover}.

\SFIG{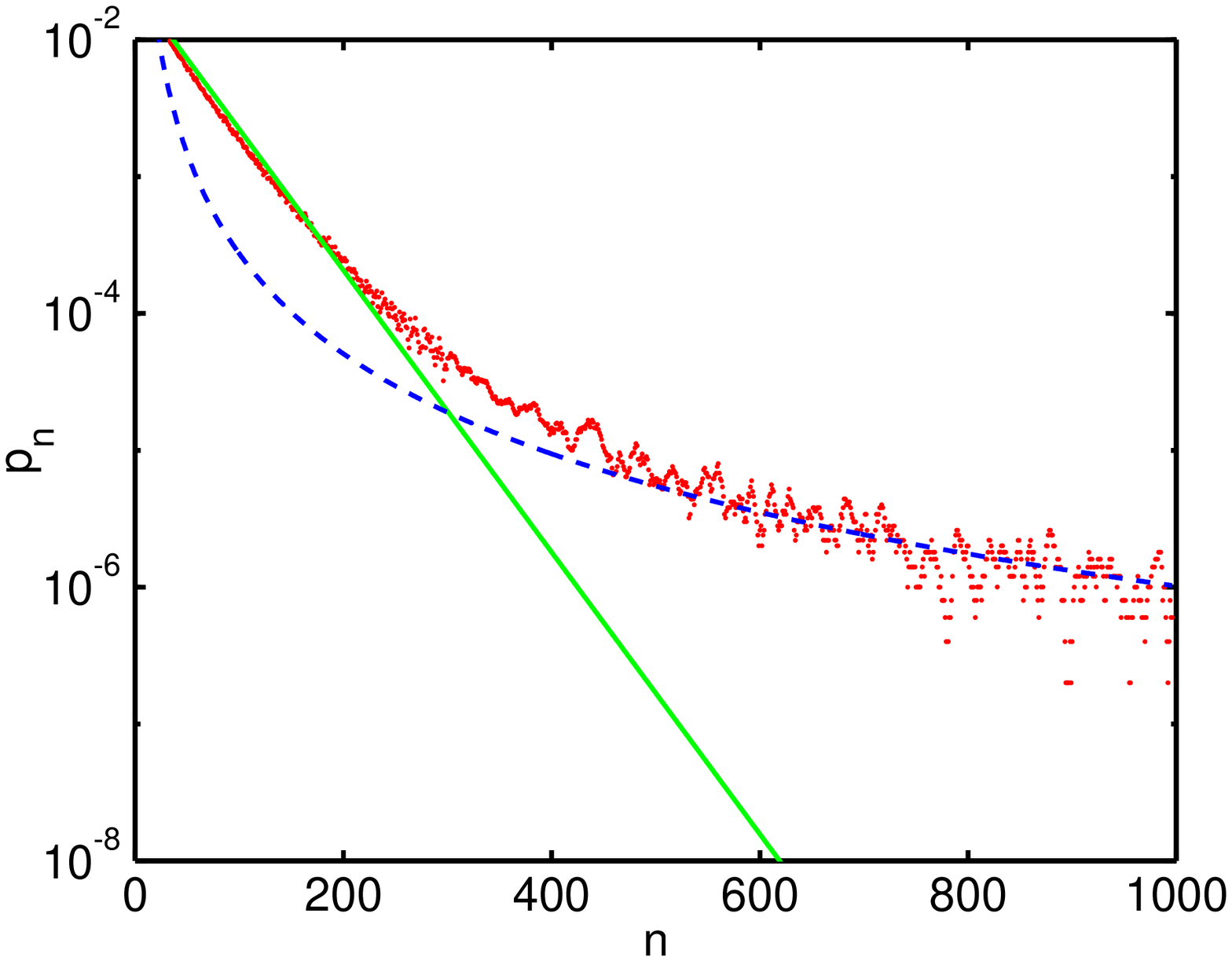}
{}{
The escape from an intermittent repeller
typically starts out exponentially,
controlled by the zeros
close to the cut but beyond the branch point $z=1$,
as in \reffig{finter_cont1},
followed by
a power law decay controlled by the type of the branch cut.
}{finter_cover}

For general, non-integer $1/s>0$, we write
\[
\zetaInv_0(z)= A(u) + (u)^{1/s} B(u)\equiv G(u)
\]
with $u = 1-z$ and
$A(u)$, $B(u)$ are functions analytic in a disc of radius 1 around
$u=0$. The leading terms in the Taylor series expansions of $A(u)$
and $B(u)$ are
\[ a_0 = \frac{b-a}{1-a}, \quad b_0 = \frac{a}{g_s(1)} \frac{1-b}{1-a},\]
see (\ref{zeta-toy-ana-gen1}).
Expanding $\frac{d}{d u} \log G(u)$ around $u=0$, one again obtains
leading order contributions according to (\ref{intereq:logG}) and
the general result follows immediately using (\ref{intereq:useful1})
\beq
\Gamma_n\sim \
\frac{a}{s g_s(1)}\frac{1-b}{b-a} \frac{1}{\Gamma(1-1/s)}
 \frac{1}{n^{1/s}}
\,.
\ee{intereq:interesc-gen1}
Applying the same arguments for integer intermittency exponents $1/s = \ell$,
one obtains
\beq
\Gamma_n\sim \ (-1)^{\ell+1}
\frac{a}{s g_{\ell}(1)}\frac{1-b}{b-a}
 \frac{\ell!}{n^{\ell}}
\,.
\ee{intereq:interesc-gen2}

So far, we have considered the survival probability for
a repeller, that is we assumed $a<b$. The formulas
(\ref{intereq:interesc-gen1}) and (\ref{intereq:interesc-gen2})
do obviously not apply for the case $a=b$ of a
bounded map. The coefficient $a_0= (b-a)/(1-a)$ in the series representation
of $G(u)$ is zero and we need to take into account next leading terms in
the expansion of the logarithmic derivative of $G(u)$ in
(\ref{intereq:logG}). One obtains
\[
\frac{d}{d u}
\log G(u)=\left\{ \begin{array}{cc}
\frac{1}{u} \left(
1+O(u^{1/s-1}) \right)   & s<1\\
\frac{1}{u} \left(
\frac{1}{s}+O(u^{1-1/s}) \right)   & s>1
\end{array}  \right.   ,
\]
where we assume $1/s$ non-integer for convenience.
The survival probability is thus in leading order
\[
\Gamma_n \sim \left\{
\begin{array}{cc}
1+O(n^{1-1/s}) & s<1\\
1/s+O(n^{1/s-1}) & s>1
\end{array}  \right.
\,.
\]
For $s<1$, this is what we expect. There is no escape, so the
survival probability is equal to 1, which we get as an asymptotic
result here.  The result for $s>1$ is somewhat more worrying. It says
that $\Gamma_n$ defined as sum over the instabilities of the periodic
orbits (\ref{escape-hyp}) does not tend to unity for large $n$.
However, the case $s>1$ is in many senses anomalous. For instance,
the invariant density cannot be normalized\cite{Hungar96}.
It is therefore not reasonable to expect that periodic orbit theories
will work without complications.

\section{Smooth intermittent maps}
\label{ss-inter-gen}

Now we turn to the problem of constructing cycle expansions
for general smooth 1-dimensional maps
with a single isolated marginal fixed point.
To keep the notation simple, we will consider two-branch maps with a complete
binary symbolic dynamics as in (\ref{int.pl}).
The necessary modifications for multi-branch maps with an isolated marginal
fixed point are straightforward.
We again assume that the behavior near the fixed point is given by
(\ref{intereq:mapprot}). This implies that the stability of a family
of cycles approaching the marginally stable orbit, for example
the family $10^m$, will increase only algebraically.
Again we find
for large $m$
\[ \frac{1}{\Lambda_{10^m}} \sim \frac{1}{m^{1+1/s}}\ , \]
where $s$ denotes the intermittency exponent.

Zeta functions or \fd s in terms of products over cycles
are formally identically to those of hyperbolic systems
except for the omission of the marginal orbit $0$, see (\ref{zetaind-def}).
When considering cycle expansion as in \refsect{sec:PF-hyp},
periodic orbit contributions from, for example,
cycles with itinerary $10^m$  are unbalanced
and we arrive at a cycle expansion in terms of infinitely many fundamental
terms\cite{AACII,Tan96} as for our toy map. This corresponds to moving from our
binary symbolic
dynamics to an infinite symbolic dynamics by making the identification
\[ 10^{m-1} \to m;\quad  10^{k-1}10^{m-1} \to km; \quad
10^{l-1}10^{k-1}10^{m-1}\to lkm; \ldots
\,,
\]
\reftab{tab:int-inf-alf}. The topological length of the orbit is
thus no longer determined by the iterations of the two-branch map $f$, but
by the number of times the cycle reaches the right branch,
that is, by the number
of symbols $1$ in the itinerary, or by the number of letters in the
symbolic dynamics of the induced map $\hat{f}$ of $f$.

\begin{table}
\begin{center}
\begin{tabular}{|rrrrrrr|}\hline
\multicolumn{2}{|c}{$\infty$ -- alphabet}&
\multicolumn{5}{c|}{binary alphabet}\\ \hline
& &$m$ = 1&$m$ = 2&$m$ = 3&$m$ = 4&$m$ = 5 \\ \hline\hline
1-cycle&$m$   &       1&      10&     100&    1000& 10000\\
2-cycle&$km$ &        &        &        &        & \\
&$1m$  &      11&     110&    1100&   11000&   110000\\
&$2m$  &     101&    0101&   10100&  101000&  1010000\\
&$3m$  &    1001&   10010&  100100& 1001000& 10010000\\
&$4m$  &   10001&  100010& 1000100&10001000&100010000\\
3-cycle&$lkm$&   &        &         &          &        \\
&$11m$&     111&    1110&    11100&    111000&    1110000\\
&$12m$&    1101&   11010&   110100&   1101000&   11010000\\
&$13m$&   11001&  110010&  1100100&  11001000&  110010000\\
&$21m$&    1011&   10110&   101100&   1011000&   10110000\\
&$22m$&   10101&  101010&  1010100&  10101000&  101010000\\
&$23m$&  101001& 1010010& 10100100& 101001000& 1010010000\\
&$31m$&   10011&  100110&  1001100&  10011000&  100110000\\
&$32m$&  100101& 1001010& 10010100& 100101000& 1001010000\\
&$33m$& 1001001&10010010&100100100&1001001000&10010010000\\ \hline
\end{tabular}
\end{center}
\vspace{0.5cm}
\caption[soso]{Infinite alphabet versus the original binary alphabet
for the shortest periodic orbit families.}
\label{tab:int-inf-alf}
\end{table}
For generic intermittent maps, curvature contributions in the cycle
expanded zeta function will not vanish exactly as they do for our
piecewise linear model. The most natural way to organize the cycle
expansion in this case is to collect cycles and pseudo-cycles
of the same topological length with respect to the infinite
alphabet of the induced map. Denoting cycle weights in the new alphabet as
$t_{km\ldots} = t_{10^{k-1}10^{m-1}\ldots}$, one obtains\cite{Tan96}
\begin{eqnarray} \label{int:cyc-exp-inf}
\zeta_0^{-1} & =& \prod_{p\ne 0} \left(1 - t_p \right)
= 1 - \sum_{n = 1}^{\infty} c_{n} \\
  &=& 1 - \sum_{m=1}^{\infty} t_m
 - \sum_{k=1}^{\infty} \sum_{m=1}^{\infty} \frac{1}{2} (t_{km} - t_k t_m)
   \nonumber\\
  &-&  \sum_{l=1}^{\infty} \sum_{k=1}^{\infty} \sum_{m=1}^{\infty}
        \left( \frac{1}{3} t_{lkm} - \frac{1}{2}t_{lk}t_m
    + \frac{1}{6} t_l t_k t_m\right)
   - \sum_{j=1}^{\infty}\sum_{l=1}^{\infty} \sum_{k=1}^{\infty}
        \sum_{m=1}^{\infty} \ldots \nonumber \; .
\end{eqnarray}
with $t_p = z^{n_p}/|\Lambda_p|$ and $n_p$ is the
length of the binary itinerary.
The first sum is the fundamental term, already noted in the
cycle expansion \refeq{cyc-exp-piecelin} for the toy model.
The curvature terms $c_n$
for $n \ge 2$ are now $n$-fold infinite sums where the prefactors take
care of double counting of prime cycles.

We consider the fundamental term first. For generic intermittent
maps, we can not expect to obtain an analytic expression for infinite
sums of the form
\beq
f(z)=\sum_{n=0}^{\infty}h_n z^n
\ee{int:power-series}
with algebraically decreasing coefficients
$h_n\, \sim n^{-(1/s+1)}$.

To evaluate the sum, we face the same problem as for our toy map: the power
series diverges for $|z|> 1$, that is, exactly in the `interesting' region
where poles, zeros or branch cuts of the zeta function are to be expected.
By carefully subtracting the asymptotic behavior with the help of
\refeq{bra-int}, one can in general construct
an analytic continuation of $f(z)$ around $z = 1$ of the form\cite{PDresum}

\begin{eqnarray}\label{int:power-series-ana}
f(z) &\sim& A(z) + B(z) (1-z)^{1/s} \qquad
                           1/s \notin \naturals \\
f(z) &\sim& A(z)+ B(z) (1-z)^{1/s}\ln (1-z) \qquad 1/s \in \naturals
\,,
\nonumber
\end{eqnarray}
where $A(z)$ and $B(z)$ are functions analytic in a disc around $z = 1$.
From here we can derive
the asymptotic behavior which coincides with the asymptotics obtained in
(\ref{intereq:interesc-gen1}) and (\ref{intereq:interesc-gen2}), that is,
the survival probability is in general given as
\beq
 \Gamma_n \sim \frac{1}{n^{1/s}} \quad \mbox{for} \quad n\to\infty
\ee{int:esc-gen-map}
for 1-dimensional maps of the form (\ref{intereq:mapprot}).
We have to work a bit harder if we want more detailed information like
exponential precursors given by zeros or poles
of the \dzeta\ or higher order corrections. This information is buried
in the functions $A(z)$ and $B(z)$ or more generally in the analytically
continued zeta functions including curvature contributions. To get such an
analytic continuation, one may follow two different strategies
which we will sketch next.

\subsection{Resummation}
\label{ss-inter-resum}

One way to get information about the zeta function
near the branch cut
is to derive the leading coefficients in the Taylor series of the
functions $A(z)$ and $B(z)$ in (\ref{int:power-series-ana}) at $z = 1$. This can
be done in principle, if the coefficients $h_n$ in sums like
(\ref{int:power-series}) are known (as for our toy model).
One then considers a resummation of the
form\cite{PDresum,PredPD,PDescape,intFellcite}
\beq
\sum_{j=0}^{\infty} h_j z^j=\sum_{j=0}^{\infty} a_j (1-z)^j+
(1-z)^{1/s} \sum_{j=0}^{\infty} b_j
(1-z)^j,
\ee{intereq:cab}
and the coefficients $a_j$ and $b_j$ are obtained in terms of the
$h_j$'s by expanding $(1-z)^{j}$ and $(1-z)^{j+1/s}$
around $z =0$ using \refeq{bra-int},
and equating the coefficients.

In practical calculations one often has only a finite number of
coefficients $h_j$, $0\leq j\leq n_\nCutoff$, which may have been obtained
by finding periodic orbits and their stabilities numerically.
One can still design a resummation scheme for the computation of the
coefficients $a_j$ and $b_j$ in (\ref{intereq:cab}).  We replace the
infinite sums in (\ref{intereq:cab}) by finite sums
of increasing degrees $n_a$ and $n_b$, and require that
\begin{equation}
\sum_{i=0}^{n_a} a_i (1-z)^i+
(1-z)^{1/s} \sum_{i=0}^{n_b} b_i
(1-z)^i=\sum_{i=0}^{n_\nCutoff} h_i z^i+O(z^{n_\nCutoff+1})
  \label{intereq:genserf} \ \ .
\end{equation}
One proceeds again by expanding the right hand side
around $z=0$, skipping all powers $z^{n_\nCutoff+1}$ and higher,
and then equating coefficients.
It is natural to require that $|n_b +1/s -n_a|<1$, so that the maximal
powers of the two sums in (\ref{intereq:genserf}) are adjacent.
If one chooses
$n_a +n_b+2=n_\nCutoff+1$, then, for each cutoff length $n_\nCutoff$,
the integers $n_a$ and $n_b$ are uniquely
determined from a linear system of equations. The prize we pay is that the
so obtained coefficients depend on the cutoff $n_\nCutoff$.
One can now study convergence of the coefficients $a_j,$ and $b_j,$
with respect to increasing values of $n_\nCutoff$,
or various quantities  derived from $a_j,$ and $b_j$. Note that the leading
coefficients $a_0$ and $b_0$ determine the prefactors as in
(\ref{intereq:interesc}).
The resummed expression can also be used to compute
zeros, inside or outside the radius of convergence of
the cycle expansion $\sum h_j z^j$.

The scheme outlined in this section tacitly assumes
that a representation of the form (\ref{int:power-series-ana})
holds in a disc of radius 1 around $z = 1$. Convergence is
improved further if additional information about the
asymptotics of sums like (\ref{int:power-series})
is used in the ansatz (\ref{intereq:cab}).

\subsection{Analytical continuation by integral transformations}
\label{ss-inter-integ}

We will now introduce a method which provides an
analytic continuation of sums of the form (\ref{int:power-series})
without relying on the ansatz (\ref{intereq:cab})\cite{Tan95,Tan96}.
The main idea is to rewrite the sum (\ref{int:power-series})
as a sum over integrals with the help of Poisson summation
and find an analytic continuation of each integral by
contour deformation.
In order to do so, we need to know the $n$ dependence
of the coefficients $h_n \equiv h(n)$ explicitly for all $n$.
If the coefficients are not known analytically, one may proceed by
approximating the large $n$ behavior in the form
\[
h(n) = n^{-1/s-1} (C_1 + C_2 n^{-1} + \ldots)
\,,\qquad
n \neq 0
\,,
\]
and determine the constants $C_i$ numerically from periodic
orbit data. By using the Poisson resummation identity
\begin{equation}
\sum_{n=-\infty}^{\infty} \delta(x - n) = \sum_{m=-\infty}^{\infty}
\exp(2\pi { i} m x)
\,,
\end{equation}
we may write the sum
(\ref{int:power-series}) as
\begin{equation} \label{int:int-exp}
f(z) =
\frac{1}{2} h(0) + \sum_{m=-\infty}^{\infty} \int_{0}^{\infty}dx\,
e^{2\pi{ i} m x} h(x) z^x.
\end{equation}
The continuous variable $x$  corresponds to the discrete summation
index $n$ and it is convenient to write $z = r \exp({ i} \sigma)$
from now on.
The integrals are of course still not convergent for $r> 0$.
An analytic continuation can be found by considering the contour
integral, where the contour goes out along the real axis, makes a
quarter circle to either the positive or negative imaginary axis and goes back
to zero.  By letting the radius of the circle go to infinity,  we
essentially rotate the line of integration form the real onto the imaginary axis.
For the $m=0$ term in (\ref{int:int-exp}), we transform $x \to { i} x$
and the integral takes on the form
\[ \int_{0}^{\infty} dx\, h(x) \, r^x\, e^{{ i} x \sigma}
=  { i} \int_0^{\infty} dx\, h(ix)\, r^{{ i}x} e^{ - x \sigma}.\]
The integrand is now exponentially decreasing for all $r>0$ and
$\sigma \ne 0$ or $ 2 \pi$.  The last condition reminds us again of the
existence of a branch cut at $\Re z \ge 1$.  By the same technique, we
find the analytic continuation for all the other integrals in
(\ref{int:int-exp}).  The real axis is then rotated according to
$x \to \pm { i} x$ where $\pm$ sign refers to the sign of $m$,
\[
\int_{0}^{\infty} dx \,e^{\pm 2\pi{ i} |m| x} h(x) \, r^x
e^{{ i} x \sigma} =  {\rm \pm i} \int_0^{\infty} dx \,
h(\pm { i} x) \, r^{\pm { i}x} e^{ - x (2 \pi |m| \pm \sigma)}.
\]
Changing summation and integration, we can carry out the sum over
$ |m|$ explicitly and one finally obtains the standard formula
\begin{eqnarray} \label{int:int_ana}
f(z) &=& \frac{1}{2} h(0)
+  { i} \int_0^{\infty} dx\, h(ix)\, r^{{ i}x} e^{ - x \sigma} \\
       &+& { i} \int_{0}^{\infty} dx\, \frac{e^{-2\pi x}}{1 - e^{-2\pi x}}
       \left[ h({ i} x)  r^{{ i}x} e^{ - x \sigma}
        - h(-{ i} x)  r^{-{ i}x} e^{ x \sigma} \right]. \nonumber
\end{eqnarray}
The transformation from the original sum to the two integrals in
(\ref{int:int_ana}) is exact for $r \leq 1$, and provides
an analytic continuation for $r >0$. The expression (\ref{int:int_ana})
is especially useful for an efficient numerical calculations of the \dzeta\
for $|z|>1$, which is essential when searching for zeros and poles of
the zeta function.

\subsection{Curvature contributions}
\label{ss-inter-curv}
So far, we have discussed the fundamental term
$\sum_{n=1}^{\infty} t_n$ in (\ref{int:cyc-exp-inf})
and showed ways to extract the
information from such
power series with algebraically decreasing coefficients. The fundamental
term determines the main structure of the zeta function in terms of the
leading order branch cut. Corrections to both the zeros and poles
of the \dzeta\ as
well as the leading and sub-leading order terms in expansions like
(\ref{int:power-series-ana}) are contained in the curvature terms in
(\ref{int:cyc-exp-inf}). The first curvature correction
\[
\sum_{k=1}^{\infty} \sum_{m=1}^{\infty} \frac{1}{2} (t_{km} - t_k t_m)
\]
also has algebraically decaying coefficients which diverge
for $|z| > 1$. The analytically continued curvature terms have
again branch cuts along the positive real $z$ - axis.
Our ability to calculate the higher order
curvature terms depends on how much we know about the
cycle weights $t_{km}$. The cycle expansion itself suggests that
the terms $t_{km}$ decrease asymptotically
like
\beq t_{km} \sim \frac{1}{(k m)^{1+1/s}} \ee{cyc-curv-asym-2}
for 2-cycles of the induced map and in general for $n$-cycles like
\[ t_{m_1m_2\ldots m_n} \sim \frac{1}{(m_1m_2\ldots m_n)^{1+1/s}}. \]
If we know the cycle weights $t_{m_1m_2\ldots m_n}$ analytically, we may
proceed like in \refsect{ss-inter-integ} by transforming the multiple sums into
multiple integrals and by rotating the axis' of integration.

\section{Summary and conclusions}

Marginally stable orbits affect
the analytic structure of \dzeta s
and the rules for constructing cycle expansions:
While the marginal orbits have to be omitted, the
cycle expansions need to include families of infinitely
many  longer
and longer unstable orbits which accumulate towards the marginally stable
cycles.
Correlations for such non-hyperbolic systems decay
algebraically with the decay rates controlled by
the branch cuts of \dzeta s.
Compared to pure hyperbolic systems,
the physical consequences are drastic: exponential decays are replaced
by slow power law  decays, and transport properties, such as the diffusion,
may become anomalous\cite{intACL,DasBuchAnDiff}.

\section*{Acknowledgements}
The authors are grateful to P.~Dahlqvist for his
key contributions
to this work and to the ``Intermittency'' chapter of
ref.~\citen{DasBuch},
and to  T. Prellberg for critical comments.
G.T. and R.A.  thank the Center for Nonlinear Science
for hospitality at Georgia Tech, where part of the work was done
with support by Glen~P. Robinson Chair.
R.A. was partially supported by the PRIN-2000 project ``Chaos and
localization in classical and quantum systems", by INFM PA ``Weak
chaos: theory and applications'', and
by EU contract QTRANS Network (``Quantum transport on an atomic scale'').
G.T. acknowledges support by the Nuffield foundation, the
EPSRC and the Royal Society.

\end{document}

%% file: def.tex
\makeatletter

\newcommand{\refeq}  [1] {(\ref{#1})}

\newcommand{\reffig} [1] {fig.~\ref{#1}}

\newcommand{\reftab} [1] {table~\ref{#1}}

\newcommand{\refsect}[1] {sect.~\ref{#1}}

\newcommand{\beq}{\begin{equation}}
\newcommand{\continue}{\nonumber \\ }
\newcommand{\nnu}{\nonumber}
\newcommand{\eeq}{\end{equation}}
\newcommand{\ee}[1] {\label{#1} \end{equation}}
\newcommand{\bea}{\begin{eqnarray}}

\newcommand{\eea}{\end{eqnarray}}
\newcommand{\barr}{\begin{array}}
\newcommand{\earr}{\end{array}}

\newcommand{\BFIG}[5]{\begin{figure}
		      \hspace*{0.10\textwidth}%
		      \begin{minipage}[b]{1.00\textwidth}
		      \epsfxsize = #1\textwidth
		      \centering{
		      \epsfbox{Fig/#2}
 		         }
                      \caption[#3]{#4}
                      \label{#5}
		      \end{minipage}
		      \end{figure} }

\newcommand{\SFIG}[4]{\begin{figure}[!]
		      \hspace*{0.10\textwidth}
		      \begin{minipage}[b]{0.45\textwidth}
                      \caption[#2]{#3}
                      \label{#4} 
		      \end{minipage}~~~~~%
		      \begin{minipage}[b]{0.40\textwidth}
		      \epsfxsize = 1.0\textwidth
		      \epsfbox{Fig/#1}
		      \end{minipage}
		      \end{figure} }

\newcommand{\ie}{{that is}}		

\newcommand{\wwwcb}{{\tt www.nbi.dk/\-Chaos\-Book/}}

\newcommand{\evOper}{evolution oper\-ator}

\newcommand{\FPoper}{Perron-Frobenius oper\-ator} 

\newcommand{\dzeta}{dyn\-am\-ic\-al zeta func\-tion}

\newcommand{\Fd}{spec\-tral det\-er\-min\-ant}
\newcommand{\fd}{spec\-tral det\-er\-min\-ant}

\newcommand{\reals}{\mathbb{R}}
\newcommand{\complex}{\mathbb{C}}

\newcommand{\naturals}{\mathbb{N}}

\renewcommand\Re{{\rm Re\,}}
\renewcommand{\det}{\mbox{\rm det}\,}

\newcommand{\tr}{\mbox{\rm tr}\,}

\newcommand{\zfct}[1]{\zeta ^{-1}_{#1}}     
\newcommand{\zetaInv}{{1/\zeta}}

\newcommand{\BER}[1]{{\mbox{\footnotesize BER}}} 

\newcommand{\pS}{{\cal M}}			
\newcommand{\Lop}{{\cal L}}	   

\newcommand{\ExpaEig}{\Lambda}

\newcommand{\cl}[1]{{n_{#1}}}	
\newcommand{\nCutoff}{N}	

\newcommand{\MarkGraph}{Markov graph}

\newcommand {\id}{{\ \hbox{{\rm 1}\kern-.6em\hbox{\rm 1}}}}

%% file: intermit.bbl
\begin{thebibliography}{99}


\bibitem{MP79} P.~Manneville and Y.~Pomeau,
    {\em Phys. Lett. \bf 75A} (1979), 1
\bibitem{intPM}  Y.~Pomeau and P.~Manneville, {\em Commun. Math. Phys.}
{\bf 74} (1980), 189

\bibitem{ham} R.S.~MacKay and J.D.~Miess, {\em Hamiltonian Dynamical Systems}
        (Adam Hilger, Bristol 1987)


\bibitem{maye} D.H.~ Mayer,
       {\em Bull. Soc. Math. France} {\bf 104} (1976), 195

\bibitem{MaRo} D.~Mayer and G.~Roepstorff,
    {\em J. Stat. Phys.} {\bf 47} (1987), 149

\bibitem{ACK} R. Artuso, P. Cvitanovi\'c and B.G. Kenny,
{\em Phys. Rev. \bf A39} (1989), 268

\bibitem{PC87} P. Cvitanovi\'c, in
    P. Zweifel, G. Gallavotti and M. Anile, eds.,
        { \em Non-linear Evolution and Chaotic Phenomena},
        (Plenum, New York 1987),
     349-361

\bibitem{C92a} P. Cvitanovi\'c, {\em Circle maps: irrationally winding},
    in C. Itzykson, P. Moussa and M. Waldschmidt, eds.,
    {\em Number Theory and Physics, Les Houches 1989 Spring School},
    (Springer, New York 1992)

\bibitem{AACII} R. Artuso, E. Aurell and P. Cvitanovi\'c,
        {\em Nonlinearity \bf 3} (1990), 361

\bibitem{Mayer91} D. H. Mayer,
     {\em Continued fractions and related transformations}, in
     {\em Ergodic Theory, Symbolic Dynamics and
         Hyperbolic Spaces}, T. Bedford, M. Keane and C. Series, eds
         (Oxford University Press, Oxford, 1991)

\bibitem{sum_rules} P. Cvitanovi\'c, Kim Hansen, J.~Rolf and G.~Vattay,
        {\em Nonlinearity \bf 11} (1998), 1209

\bibitem{Tan95} G.~Tanner and D.~Wintgen, {\em Phys.~Rev.~Lett.} {\bf 75}
                (1995), 2928

\bibitem{Tan96} G.~Tanner, K.T.~Hansen and J.~Main, {\em Nonlinearity} {\bf 9}
                (1996), 1641

\bibitem{Tan97} G.~Tanner, {\em J.\ Phys.\ A} {\bf 30}
                (1997), 2863

\bibitem{ruelle-b} D. Ruelle,
        {\em Statistical Mechanics, Thermodynamic Formalism}
        (Addison-Wesley, Reading MA, 1978)

\bibitem{PG97} P. Gaspard,
        {\em  Chaos, Scattering and Statistical Mechanics}
        (Cambridge Univ. Press, Cambridge 1997)


\bibitem{DasBuch} P.~Cvitanovi\'c, R.~Artuso, P.~Dahlqvist, R.~Mainieri,
    G.~Tanner, G.~Vattay, N.~Whelan and A.~Wirzba,
    {Chaos: Classical and Quantum}
    ({\tt \wwwcb} 2003)

\bibitem{Rugh92} H.H. Rugh,
        {\em Nonlinearity \bf 5} (1992), 1237  

\bibitem{CRR93} P.~Cvitanovi\'c, P.E.~Rosenqvist, H.H.~Rugh, and G.~Vattay,
         {\em CHAOS \bf 3} (1993), 619 

\bibitem{DC97} C.P. Dettmann and P. Cvitanovi\'c,
        {\em Phys. Rev. \bf E 56} (1997), 6687;
    ``Stability ordering of cycle expansions'' section
    of ref.~\citen{DasBuch}

\bibitem{Dah94} P.~Dahlqvist,
    {\em J.\ Phys.\ A}  {\bf 27} (1994), 763

\bibitem{Dah95} P.~Dahlqvist,
    {\em Nonlinearity}  {\bf 8} (1995), 11

\bibitem{PC89} P. Cvitanovi\'c,
    {\em Phys. Rev. Lett. \bf 61} (1988), 2729

\bibitem{AACI} R. Artuso, E. Aurell and P. Cvitanovi\'c,
        {\em Nonlinearity \bf 3} (1990), 325

\bibitem{intRugh} H.H. Rugh, {\em Inv. Math. \bf 135} (1999), 1

\bibitem{intIso} S. Isola,
    {\em J. Stat. Phys. \bf 97} (1999), 263

\bibitem{SIF} S. Isola, {\em Nonlinearity \bf 15} (2002), 1521

\bibitem{TPF} T. Prellberg,
    {\em Complete Determination of the Spectrum of a Transfer Operator associated with Intermittency},
    {\tt nlin.CD/0108044}

\bibitem{intPrell} T. Prellberg, {\em Maps of the interval with indifferent
fixed points: thermodynamic formalism and phase transitions}, Ph.D. Thesis
(Virginia Polytechnic Inst. 1991).
\bibitem{PS92} T. Prellberg and J. Slawny,
        {\em J. Stat. Phys. \bf 66} (1992), 503 

\bibitem{intGasWan} P. Gaspard and X.-J. Wang,
    {\em Proc. Natl. Acad. Sci. U.S.A. \bf 85} (1988), 4591;
    X.-J. Wang, {\em Phys. Rev.} {\bf A40} (1989), 6647;
    X.-J. Wang, {\em Phys. Rev.} {\bf A39} (1989), 3214

\bibitem{FK} B. Fornberg and K.S. K\"olbig,
       {\em Math. of Computation \bf 29} (1975), 582
\bibitem{Htf1} A. Erd\'elyi, W. Magnus, F. Oberhettinger and F. G. Tricomi,
    {\em Higher trancendental functions, Vol. I}
    (McGraw-Hill, New York, 1953).


\bibitem{PDalhlqProof} The proof of the bound
\refeq{intereq:bound1weak} can be found  in P.~Dahlqvist's
unpublished notes, {\tt  {\wwwcb}extras/PDahlqvistEscape.ps.gz}.

\bibitem{Hungar96} Z. Kaufmann, H. Lustfeld, and J. Bene,
    {\em Phys. Rev. \bf E 53} (1996), 1416

\bibitem{PDresum} P.~Dahlqvist,
    {\em J.\ Phys.\ A\/ \bf 30} (1997), L351
\bibitem{PredPD}
        S.F.~Nielsen, P.~Dahlqvist, P.~Cvitanovi{\a'{c}},
        {\em J. Phys. \bf A 32} (1999), 6757  

\bibitem{PDescape} P.~Dahlqvist,
    {\em Phys.\ Rev.\ E\/ \bf 60} (1999), 6639
\bibitem{intFellcite} for an introduction to Tauberian theorems for power series
            and Laplace transforms see\\
         W. Feller, {\em An introduction to probability theory and
         applications, Vol. II} (Wiley, New York 1966).


\bibitem{intACL} R. Artuso, G. Casati and R. Lombardi,
    {\em Phys. Rev. Lett. \bf 71} (1993), 62
\bibitem{DasBuchAnDiff}
    Section {\em Marginal stability and anomalous diffusion},
    ref.~\citen{DasBuch}.

\end{thebibliography}
